\documentclass{ws-ijmpe}

\def\kt{{k_{_\perp}}}

\begin{document}

\markboth{R. Perez-Ramos}{Light and heavy quark inclusive hadronic distributions}

\catchline{}{}{}{}{}

\title{The internal structure of jets at colliders:
light and heavy quark inclusive hadronic distributions
}

\author{\footnotesize Redamy Perez-Ramos}

\address{Departament de F\'isica Te\`orica and IFIC, Universitat de Val\`encia-CSIC\\
Dr. Moliner 50, Burjassot E-46100,
Spain\\
redamy.perez@uv.es}

\maketitle

\begin{history}
\received{(received date)}
\revised{(revised date)}
\end{history}

\begin{abstract}
In this paper, we report our results on charged hadron
multiplicities of heavy quark initiated jets produced in high 
energy collisions. After implementing the so-called dead cone effect in QCD 
evolution equations, we find that the
average multiplicity decreases significantly as compared to the massless case. 
Finally, we discuss on the transverse momentum distribution of light quark initiated 
jets and emphasize on the comparison between our predictions and CDF data.
\end{abstract}

\section{Introduction}

High-$p_t$ jets can be initiated either in a short-distance
interaction among partons in high energy collisions such as $pp$, 
$p\bar p$, in the DIS $e^\pm p$, the $e^+e^-$ annihilation and via
electroweak (or new physics) processes.

In this framework, we compute the average (charged) multiplicity 
of a jet initiated by a heavy quark. For this purpose,
we extend the modified leading logarithmic approximation (MLLA) 
evolution equations \cite{Dokshitzer:1991wu} to the case where the 
jet is initiated by a heavy (charm, bottom) quark. 
The average multiplicity of light quark jets produced in high energy collisions can be 
written as $N(Y)\propto\exp\left\{\int^Y\gamma(y)dy\right\}$ ($Y$ and $y$ are
defined in section \ref{sec:kivar}), where $\gamma\simeq\gamma_0+\Delta\gamma$ 
is the anomalous dimension that accounts for
soft and collinear gluons in the double logarithmic 
approximation (DLA) $\gamma_0\simeq\sqrt{\alpha_s}$, in addition to
hard collinear gluons $\Delta\gamma\simeq\alpha_s$ or single logarithms (SLs),
which better account for energy conservation and the running of the 
coupling constant $\alpha_s$ \cite{Dokshitzer:1991wu}.  

The inclusive $k_\perp$-distribution of particles
inside a jet has been computed at MLLA accuracy
in the limiting spectrum
approximation~\cite{PerezRamos:2005nh}, {\it i.e.} assuming an infrared cutoff 
$Q_0$ equal to $\Lambda_{QCD}$ (for a review, see also \cite{Dokshitzer:1991wu}).
MLLA corrections, of relative magnitude ${\cal O}{(\sqrt{\alpha_s})}$ with respect
to the leading double logarithmic approximation (DLA), were shown to be quite
substantial ~\cite{PerezRamos:2005nh}. Therefore, we also included
corrections of order ${\cal O}{(\alpha_s)}$, that is
next-to-next-to-leading or Next-to-MLLA (NMLLA).

Lastly, both observables, the average multiplicity and the inclusive
$k_\perp$-distribution are computed under the assumption of local parton 
hadron duality (LPHD) as hadronization model 
\cite{Azimov:1984np,Dokshitzer:1995ev}.

\section{Kinematics and variables}
\label{sec:kivar}
As known from jet calculus for light quarks, 
the evolution time parameter determining the structure
of the parton branching of the primary gluon is given by 
(for a review see \cite{Dokshitzer:1991wu} and references therein)
\begin{equation}\label{eq:masslessev}
y=\ln\left(\frac{k_\perp}{Q_0}\right),\quad k_\perp=zQ\geq Q_0,
\quad Q=E\Theta\geq Q_0,
\end{equation}
where $k_\perp$ is the transverse momentum of the gluon emitted off the
light quark, $Q$ is the virtuality of the jet (or jet hardness), 
$E$ the energy of the 
leading parton, $Q_0/E\leq\Theta\leq\Theta_0$  
is the emission angle of the gluon ($\Theta\ll1$), $\Theta_0$ the 
total half opening angle of the jet being 
fixed by experimental requirements.
Let us define in this context the variable $Y$ as $y=Y+\ln z,\; Y=\ln\left(\frac{Q}{Q_0}\right)$.
The appearance of this scale is a consequence of angular ordering
(AO) of successive
parton branchings in QCD cascades \cite{Dokshitzer:1991wu,Azimov:1984np}. 

The corresponding evolution time parameter for a jet initiated 
by a heavy quark with energy $E$ and mass $m$ appears in a natural 
way and reads $\tilde y=\ln\left(\frac{\kappa_\perp}{Q_0}\right),\;
\kappa_\perp^2=k_\perp^2+z^2m^2$ \cite{Dokshitzer:2005ri},
which for collinear emissions $\Theta\ll1$ can also be rewritten in the form
\begin{equation}\label{eq:Qtheta}
\kappa_\perp=z\tilde Q,\quad \tilde Q=E\left(\Theta^2+\Theta_m^2\right)^{\frac12},
\end{equation}
with $\Theta\geq\Theta_m$ (see Fig.\,\ref{fig:Qsplit}). 
An additional comment is in order concerning the 
AO for gluons emitted off the heavy quark. In (\ref{eq:Qtheta}), $\Theta$ is the  
emission angle of the primary gluon $g$ being emitted off the heavy quark. Now
let $\Theta'$ be the emission angle of a second gluon $g'$ relative to the primary 
gluon with energy $\omega'\ll\omega$ and $\Theta''$ the emission angle relative to 
the heavy quark; in this case 
the {\em incoherence} condition $\Theta'^2\leq(\Theta^2+\Theta_m^2)$ together with 
$\Theta''>\Theta_m$ (the emission angle of the second gluon should still be larger 
than the dead cone) naturally leads (\ref{eq:Qtheta}) 
to become the proper evolution
parameter for the gluon subjet (for more details see \cite{Dokshitzer:2005ri}). 
For $\Theta_m=0$, the standard 
AO ($\Theta'\leq\Theta$) is recovered.
Therefore, for a massless quark, the virtuality of the
jet simply reduces to $Q=E\Theta$ as given above.
The same quantity $\kappa_\perp$ determines the scale of the running coupling
$\alpha_s$ in the gluon emission off the heavy quark. It can be related to the anomalous
dimension of the process by
\begin{equation}
\gamma_0^2(\kappa_\perp)=
2N_c\frac{\alpha_s(\kappa_\perp)}{\pi}=\frac1{\beta_0(\tilde y+\lambda)},\;
\beta_0(n_f)=\frac1{4N_c}\left(\frac{11}3N_c-\frac23n_f\right),\; 
\lambda=\ln\frac{Q_0}{\Lambda_{QCD}},
\end{equation}
where $n_f$ is the number of active flavours and $N_c$ the number of colours.
The variation of the effective 
coupling $\alpha_s$ as $n_f\to n_f+1$ over the heavy quarks threshold has 
been suggested by next-to-leading (NLO) calculations in the $\overline{MS}$ scheme 
\cite{Dokshitzer:1995ev} and is sub-leading in this frame. 
In this context $\beta_0(n_f)$ will be evaluated at the total number of 
quarks we consider in our application. The four scales of the process are related as follows,
$$
\tilde Q\gg m\gg Q_0\sim\Lambda_{QCD},
$$
where $Q_0\sim\Lambda_{QCD}$ corresponds to the limiting spectrum 
approximation \cite{Dokshitzer:1991wu}. Finally, the dead cone 
phenomenon imposes the following bounds of integration to the perturbative regime
\begin{equation}\label{eq:bounds}
\frac{m}{\tilde Q}\leq z\leq 1-\frac{m}{\tilde Q},\quad
m^2\leq\tilde Q^2\leq E^2(\Theta_0^2+\Theta_m^2),
\end{equation} 
which now account for the phase-space of the heavy quark jet. The last inequality
states that the minimal transverse momentum of the jet $\tilde Q=E\Theta_m=m$ 
is given by the mass of the heavy quark, which enters the game as the natural 
cut-off parameter of the perturbative approach.

\section{QCD evolution equations}
\begin{figure}[h]
\begin{center}
\epsfig{file=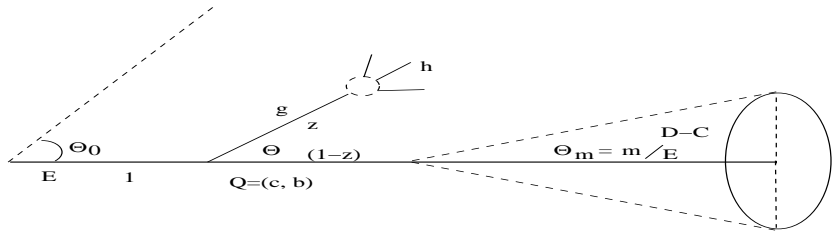, height=3truecm,width=11truecm}
\caption{\label{fig:Qsplit} Parton splitting in the process $Q\to Qg$: 
a {\em dead cone} with opening angle $\Theta_m$
is schematically shown.}  
\end{center}
\end{figure}

The system of QCD evolution 
equations for the heavy quark initiated jet, corresponding
to the process described in Fig.\,\ref{fig:Qsplit} is found to read \cite{Ramos:2010cm}
\begin{eqnarray}
\frac{N_c}{C_F}\frac{dN_Q}{d\tilde Y}&=&
\epsilon_1(\tilde Y,L_m)\int_{\tilde Y_{m}}^{\tilde Y_{ev}}
d\tilde y\gamma_0^2(\tilde y)N_g(\tilde y)-\epsilon_2(\tilde Y,L_m)\gamma_0^2(\tilde Y)N_g(\tilde Y),
\label{eq:NQhter}\\
\frac{dN_g}{d\tilde Y}&=&
\int_{\tilde Y_{m}}^{\tilde Y_{ev}}d\tilde y\gamma_0^2(\tilde y)N_g(\tilde y)-A(\tilde Y,L_m)
\gamma_0^2(\tilde Y)N_g(\tilde Y),
\label{eq:NGhter}
\end{eqnarray}
where $A(\tilde Y,L_m)$, $a(n_f)$, $\epsilon_1(\tilde Y,L_m)$ and $\epsilon_2(\tilde Y,L_m)$
are defined in \cite{Ramos:2010cm}.
As a consistency check, upon integration over $\tilde Y$ of
the DLA term in Eq.(\ref{eq:NQhter}), the phase space structure of 
the radiated quanta reads
\begin{equation}\label{eq:dconesimple}
N_Q(\ln\tilde Q)\approx1+\frac{C_F}{N_c}\int_{0}^{\Theta^2_0}
\frac{\Theta^2d\Theta^2}{(\Theta^2+\Theta_m^2)^2}\int_{m/\tilde Q}^{1-{m/\tilde Q}}\frac{dz}{z}
\left[\gamma_0^2N_g\right](\ln z\tilde Q).
\end{equation}
Notice that the lower bound over $\Theta^2$ in (\ref{eq:dconesimple}) ($\tilde Y$ in (\ref{eq:NQhter}))
can be taken down to ``0" ($Y_m=L_m$ in (\ref{eq:NQhter})) because the heavy quark mass plays the 
role of collinear cut-off parameter.
\section{Phenomenological consequences}

Working out the structure of (\ref{eq:NQhter}) and (\ref{eq:NGhter}), we obtain
the rough difference between the light and heavy quark jet multiplicities, which yields,
\begin{equation}\label{eq:Nqdiff}
N_q-N_Q\stackrel{E\to\infty}{\approx}\left[1-\exp\left(-2\sqrt{\frac{L_m}{\beta_0}}\right)\right]N_q,\quad
N_q\propto\exp2\sqrt{\frac{\tilde Y}{\beta_0}}.
\end{equation}
It can be seen that (\ref{eq:Nqdiff}) is exponentially increasing because it is dominated by
the leading DLA energy dependence of $N_q$. According to (\ref{eq:Nqdiff}), the gap 
arising from the dead cone effect should be bigger for the $b$ than for the $c$ quark
at the primary state bremsstrahlung radiation off the heavy quark jet.
The approximated solution of the 
evolution equations leads to the rough behaviour of $N_q-N_Q$ in 
(\ref{eq:Nqdiff}), which is not exact in its present form. That is why, in 
the following, we use the numerical solution of the equations (\ref{eq:NQhter}) and 
(\ref{eq:NGhter}).
\begin{figure}[h]
\begin{center}
\epsfig{file=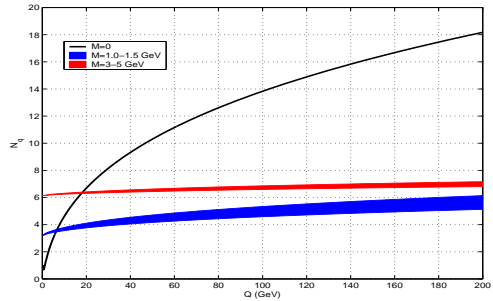, height=4truecm,width=6.5truecm}
\caption{\label{fig:NQ} Massless 
and massive quark jet average multiplicity $N_Q^{total}$ including heavy quark flavour decays 
as a function of the jet hardness $Q$. Bands indicate $m_c$ and $m_b$ in the $[1,1.5]$ and 
$[3,5]$ GeV intervals respectively.}  
\end{center}
\end{figure}
In this study we advocate the role of 
mean multiplicities of jets as a
potentially useful signature for $b$-tagging and 
new associated physics \cite{SanchisLozano:2008te} 
when combined with other selection criteria.
In Fig.\,\ref{fig:NQ}, we plot as function of the jet 
hardness, the total average jet multiplicity ($N_A^{total}$), which accounts for
the primary state radiation off the heavy quark ($N^{ch}_A$)
together with the decay products from the final-state flavoured hadrons ($N_A^{dc}$), in the form
$N_A^{total}=N^{ch}_A+N_A^{dc}$. Moreover, the flavour decays 
constants $N_c^{dc}=2.60 \pm 0.15$ and $N_b^{dc}=5.55 \pm 0.09$ \cite{Akers:1995ww} are independent
of the hard process inside the cascade, such that $N_A^{dc}$ can be added in the 
whole energy range. After accounting for the weak decay multiplicities, it turns out that the $b$ quark jet 
multiplicity becomes slightly higher than the $c$ quark jet multiplicity, although
both remain suppressed because of the dead cone effect. 

\section{Single inclusive $\boldsymbol{k_\perp}-$distribution of charged hadrons in NMLLA}
\begin{figure}
\begin{center}
\includegraphics[height=9cm,width=9.8cm]{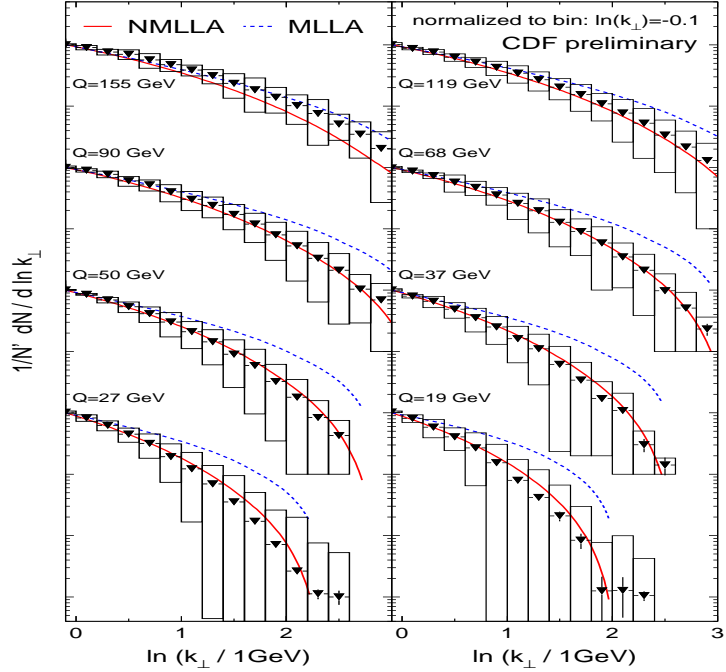}
\caption{\label{fig:CDF-NMLLA}CDF preliminary results for the inclusive
$\kt$ distribution at various hardness $Q$ in comparison to MLLA and
NMLLA predictions at the limiting spectrum ($Q_0=\Lambda_{QCD}$); the boxes are
the systematic errors.}
\end{center}
\end{figure}

Computing the single inclusive $k_\perp-$ distribution requires
the definition of the jet axis. The starting 
point of our approach consists in considering
the correlation between two particles
(h1) and (h2) of energies $E_1$ and $E_2$ which form a
relative angle $\Theta$ inside one jet of total 
opening angle $\Theta_0>\Theta$ \cite{PerezRamos:2005nh}. 
Weighting over the energy $E_2$ of particle (h2), 
this relation leads to the correlation
between the particle (h=h1)
and the energy flux, which we identify
with the jet axis \cite{PerezRamos:2005nh}. Thus,
the correlation and the relative transverse momentum 
$k_\perp$ between (h1) and (h2) are replaced by the correlation,
and transverse momentum of (h1) with
respect to the direction of the energy flux. 
In Fig.~\ref{fig:CDF-NMLLA} \cite{Arleo:2007wn}, we report measurements  over a wide
range of jet hardness, $Q=E\Theta_0$, in $p\bar{p}$ collisions at
$\sqrt{s}=1.96$~TeV~\cite{Aaltonen:2008yn}, together with the MLLA predictions of
\cite{PerezRamos:2005nh} and the NMLLA calculations, both
at the limiting spectrum ($\lambda=0$) and taking $\Lambda_{QCD}=250$~MeV;
the experimental distributions suffering from large normalization errors,
data and theory are normalized to the same bin, $\ln(\kt/1\,\text{GeV})=-0.1$.
The hadronic $k_\perp$-spectrum inside a high energy
jet is determined including corrections of relative magnitude
${\cal O}(\sqrt{\alpha_s})$ with respect to the MLLA,  in the limiting spectrum approximation.
The results in the limiting spectrum approximation are found to be in
impressive agreement with measurements by the CDF collaboration, unlike
what occurs at MLLA, pointing out small overall
non-perturbative contributions.
The MLLA predictions prove reliable in a much smaller
$\kt$ interval. At fixed jet hardness (and thus $Y_{\Theta_0}$),
NMLLA calculations prove
accordingly trust-able in a much larger $x$ interval.

\section{Conclusions}

Thus, our present work focusing on the differences of the average charged
hadron multiplicity between jets initiated by gluons, light or heavy quarks
could indeed represent a helpful auxiliary criterion to tag 
such heavy flavours from background for jet hardness $Q \gtrsim 40$ GeV.
Notice that we are suggesting as a potential signature
the {\em a posteriori} comparison
between average jet multiplicities corresponding to 
different samples of events where other criteria
to discriminate heavy from light quark initiated jets
were first applied. 
Fig.\ref{fig:NQ} plainly demonstrate that the separation between light quark
jets and heavy quark jets is allowed above a few tens
of GeV with the foreseen errors of the experimentally measured average 
multiplicities of jets. 

The single inclusive $k_\perp$-spectra inside a jet is
determined including higher-order ${\cal O}{(\alpha_s)}$ (i.e. NMLLA) corrections
from the Taylor expansion of the MLLA evolution equations and beyond the
limiting spectrum, $\lambda\ne 0$. The agreement between NMLLA predictions
and CDF preliminary data in $p\bar{p}$ collisions at the Tevatron is
impressive, indicating very small overall
non-perturbative corrections and giving further support
to LPHD \cite{Azimov:1984np}.

\end{document}